\documentclass[12pt]{article}
\usepackage{epsfig}
\usepackage{amsfonts}
\usepackage{amsopn}
\usepackage{amsmath}
\usepackage{verbatim,amsthm}

\title
{
\vskip -50 pt
\begin{flushright}
\normalsize\rm NORDITA-2009-23
\end{flushright}
\vskip 20 pt
Exact solutions for $U(1)$ globally invariant membranes
}

\author{
M. Trzetrzelewski$^{a}$\thanks{e-mail: 33lewski@th.if.uj.edu.pl} \
and A. A. Zheltukhin $^{b,c,d}$\thanks{e-mail: aaz@physto.se}  \\ \\
$^a$ M. Smoluchowski Institute of Physics, Jagiellonian University, \\
Reymonta 4, 30-059 Krak\'ow, Poland  \\ \\
$^b$ Kharkov Institute of Physics and Technology, \\
1, Akademicheskaya St., Kharkov, 61108, Ukraine \\  \\
$^c$ Fysikum, AlbaNova, Stockholm University, \\
106 91, Stockholm, Sweden \\ \\
$^d$ NORDITA,  \\
Roslagstullsbacken 23, 106 91 Stockholm, Sweden
}

\date{}

\begin{document}

\maketitle

\begin{abstract}
The exact solvability problem of the nonlinear equations describing the  $U(1)$
invariant membranes is studied and the general solution for the static membrane
in $D=2N+1$-dimensional Minkowski space-time, including M-theory case $D=11$, is obtained.
The D=5 time-dependent elliptic cosine solution describing a family of contracting
tori is also found, together with the solution corresponding to a spinning torus characterized
by the presence of the critical rotation frequency $\Omega_{max}=\frac{T^{1/3}}{\sqrt{\pi}}$
expressed via the membrane tension $T$ .

\end{abstract}

\section{Introduction}

A fundamental role of p-branes in M-theory \cite{M1} attracts much of interest
to studying  the membrane ($p=2$) nonlinear equations. However, not so much is known
about their exact classical solutions
(see e.g. \cite{tucker, hoppe1, Z_0, BZ_0, hoppe2,hoppe4,UZ, hoppe6, JU1})
and a membrane realization of the string program \cite{Reb}.
Recently Hoppe in \cite{JU1} observed that 
adding the $U(1)$ global symmetry to the problem
transforms the original 3-dim membrane equations into a 2-dim string-like nonlinear problem
still capturing characteristic nonlinearities of the membrane dynamics.

Here the solvability of Hoppe's
non-linear equations  \cite{JU1}  and their  Hamiltonian formulation are  investigated
\footnote{We thank Jens Hoppe for bringing to our attention this interesting problem.}.
We prove the exact solvability of the non-linear
problem for the $U(1)$ static membrane in
$D=2N+1$-dimensional Minkowski space, including the M-theory case $D=11$,
and present the general solution. Then the time-dependent solutions for  $D=5$
are analyzed and two physically interesting solutions are found. The
first solution is given  by the family of the Jacobi elliptic cosine
functions parametrized by the discrete integer number $n$ and describing a
contracting tori. The second solution describes a spinning torus characterized
by the upper boundary $\Omega_{max}=\frac{T^{1/3}}{\sqrt{\pi}}$ for
the rotation frequency  $\Omega$, where $T$ is the membrane  tension.

\section{The membrane dynamics}

The action for a p-brane without boundaries
is given by the integral in the world-volume
parameters $\xi^{\alpha}$ ($\alpha=0,\ldots,p$)
\footnote {Here the D-dimensional Minkowski space has
the signature $\eta_{\mu\nu}=(+,-,\ldots,-)$.}
\[
S=\int \sqrt{|G|}d^{p+1}\xi, \label{1}
\]
where $G$ is determinant of the induced metric $G_{\alpha
\beta}:=\partial_{\alpha} x_{\mu}\partial_{\beta} x^{\mu}.
\label{2}$ After splitting of the embedding
$x^{\mu}=(x^0,x^i)=(t,\vec{x})$ and internal coordinates
$\xi^{\alpha}=(\tau,\sigma^r)$, the Euler-Lagrange equations and
$p+1$ primary constraints generated by $S$ take the  form \footnote{
We  have chosen  the coordinates $x^{\mu}$  to be dimensionless
(similarly $\xi^{\alpha}$)  by rescaling  $x^{\mu}  \to x^{\mu} /
\sqrt{\alpha'}$,  where $\alpha'$ parameterizes the p-brane tension
$T \sim 1/{{\alpha'}^{\frac{p+1}{2}}}$.}
\begin{equation}\label{5}
\dot{\mathcal{P}}^{\mu}=-\partial_r \sqrt{|G|}G^{r\alpha}\partial_{\alpha}x^{\mu}, \ \ \
\mathcal{P}^{\mu}=\sqrt{|G|}G^{\tau \beta}\partial_{\beta} x^{\mu},
\end{equation}
\begin{equation}
\tilde{T}_{r}:=\mathcal{P}^{\mu} \partial_{r} x_{\mu} \approx 0, \ \ \ \
\tilde{U}:=\mathcal{P}^{\mu}\mathcal{P}_{\mu}-|\det G_{rs}| \approx  0, \label{6}
\end{equation}
where $\mathcal{P}^{\mu}$ is the energy-momentum density.
Next we follow \cite{hoppe4} and use  the orthogonal gauge
\begin{eqnarray} \label{7}
\tau=x^0, \ \ \ \ G_{\tau r}= -(\dot{\vec{x}} \cdot \partial_r \vec{x})=0, \\
 g_{rs}:=\partial_r \vec{x} \cdot \partial_s \vec{x}, \ \ \ \
G_{\alpha\beta}=\left( \begin{array}{cc}
                       1- {\dot{\vec{x}}}^2 & 0    \\
                         0 & -g_{rs}
                              \end{array} \right)
                               \nonumber
\end{eqnarray}
to simplify the metric $G_{\alpha\beta}$.
The solution of the constraint $\tilde{U}$ (\ref{6})
takes the form
\begin{equation}
\mathcal{P}_0=\sqrt{\vec{\mathcal{P}}^2+g}, \ \ \ \
g=\det g_{rs} \label{9}
\end{equation}
and becomes the  Hamiltonian density $\mathcal{H}_0$ of the p-brane,
because $\dot{\mathcal{P}}_0=0$ in view of  Eqn. (\ref{5}).
Respectively the evolution of $\vec{\mathcal{P}}$
is described as follows
\begin{equation}\label{13}
\vec{\mathcal{P}}=\mathcal{P}_0 \dot{\vec{x}}=\sqrt{\frac{g}{1-\dot{\vec{x}}^2}}\dot{\vec{x}}, \ \ \ \
\dot{\vec{\mathcal{P}}}=  \partial_r \left( \frac{g}{\mathcal{P}_0} g^{rs}\partial_s \vec{x}\right)
\end{equation}
and  Eqns. (\ref{13}) yield the second order PDE for $\vec{x}$
\begin{equation}
 \ddot{\vec{x}}=\frac{1}{\mathcal{P}_0}\partial_r \left( \frac{g}{\mathcal{P}_0} g^{rs}\partial_s \vec{x}\right).
\end{equation}
The constraints $\tilde{T}_r$ take the form
$T_r:=\vec{\mathcal{P}}\partial_r \vec{x}=0 $
and are satisfied on the surface of the gauge conditions (\ref{7}).
The gauge (\ref{7}) has a residual symmetry
\begin{equation}
\tilde{\tau}=\tau, \ \ \ \ \tilde{\sigma}^r=f^r(\sigma^s) \label{diff}
\end{equation}
which allows (see \cite{Mos})
to simplify  Eqns. (\ref{13}) by choosing
the additional gauge condition  $\mathcal{H}_0(=\mathcal{P}_0)=const.=C>0 $ resulting in
\begin{eqnarray}
U:=\vec{\mathcal{P}}^2+g-C^2=0,  \nonumber \\
\dot{\vec{x}}=\frac{1}{C}\vec{\mathcal{P}}, \ \ \ \ \dot{\vec{\mathcal{P}}}
=\frac{1}{C}\partial_r (g g^{rs} \partial_s \vec{x}).
\nonumber
\end{eqnarray}
Then the equations of motion admit the canonical formulation
\[
\dot{\vec{x}}=\{H,\vec{x}\}, \ \ \ \ \dot{\vec{\mathcal{P}}}=\{H,\vec{\mathcal{P}}\}, \ \ \ \
\{\mathcal{P}_i(\sigma), x_j(\tilde{\sigma})  \}=
\delta_{ij}\delta^{(2)}(\sigma^r-\tilde{\sigma}^r)
\]
with the Hamiltonian given by
\begin{eqnarray}\label{hampbr}
H=\frac{1}{2C}\int d^p \sigma (\vec{\mathcal{P}}^2+g).
\end{eqnarray}
Next we focus our analysis on the case of membranes ($p=2$) with
additional $U(1)$ symmetry.

\subsection{$U(1)$ invariant membranes}

The 2-dim surface  $\Sigma$ of a  $U(1)$ invariant membrane
 has the group  $U(1)$ as its isometry
with the Killing vector $\frac{\partial}{\partial \sigma^2}$.
Thus, the metric tensor $g_{rs}$ on $\Sigma$
is independent of $\sigma^2$, i.e. $\partial_{\sigma^2}g_{rs}=0$.
 Our further analysis will be narrowed to the case of $U(1)$ invariant membranes without boundaries.
 The $U(1)$ membranes with boundaries are treated similarly
taking into account additional boundary terms.
The $U(1)$  membranes may be compact or non-compact. For the latter
there are no conditions on the world-volume parameter $\sigma^1$.

The $U(1)$ membrane vector $\vec{x}$  may be chosen in the form \cite{JU1}
\begin{equation} \label{anzats}
\vec{x}^T=(m_1\cos\sigma^2,m_1\sin\sigma^2,\ldots,m_N\cos\sigma^2,m_N\sin\sigma^2), \ \
m_a=m_a(\tau,\sigma^1).
\end{equation}
It belongs to the $D=2N+1$ dimensional Minkowski space and generates the
metric  $g_{rs}=\partial_r \vec{x}\partial_s \vec{x}$ independent on $\sigma^{2}$.
The geometric restrictions associated with the representation (\ref{anzats}) are clarified from the
next observation. Because any $2N$-dim vector $\vec{x}$ is fixed
by $N$ pairs of its polar coordinates,
the space vector $\vec{x}$
of  any membrane may be presented in the form
\[
\vec{x}^T(\tau,\sigma^{1}, \sigma^{2})=
(m_1\cos\theta^1,m_1\sin\theta^1,\ldots,m_N\cos\theta^N,m_N\sin\theta^N)
\]
(where $T$ means the transposition)
with $m_a=m_a(\tau,\sigma^{r}),\  \theta^a=\theta^a(\tau,\sigma^{r})$
parametrized by its world volume coordinates $(\tau,\sigma^{1}, \sigma^{2})$.
The ansatz (\ref{anzats})  is obtained  from the above representation
by equaling all
the polar angles $\theta_a$ to $\sigma^2$ and the manifestation
of the radial coordinate $m_a=m_a(\tau,\sigma^1)$ independence on the parameter $\sigma^{2}$.
It means  geometrically that
(\ref{anzats}) describes a 2-dim surface swept by global rotations of a subgroup 
$O(2)\in SO(2N)$,  parametrized by the angle $\sigma^{2}$, of a plane 
closed  ${\bold m}$-curve given by its radial
coordinates $m_a=m_a(\sigma^1)$ at any fixed moment $\tau$.
 Some of these $O(2)$ rotation subgroups
 will create their own  $U(1)$ invariant closed 2dim surfaces.
 At least, it concerns of the $O(2)$ rotations with their axices lying in the plane of 
the ${\bold m}$-curve. 
It is easy to see that the ansatz (\ref{anzats}), e.g. in  $D=2N+1=5$ 
\begin{equation}\label{anz} 
\vec{x}^T=(m_1\cos\sigma^2,m_1\sin\sigma^2,m_2\cos\sigma^2,m_2\sin\sigma^2),
\end{equation}
is created by the following one-parametric subgroup of $SO(4)$ rotations 
\begin{equation}\label{4rot}
S=\left( \begin{array}{cccc}
                       \cos \sigma^2 & -\sin \sigma^2 & 0 & 0   \\
                         \sin \sigma^2 & \cos \sigma^2 & 0 & 0   \\
                        0 & 0   &   \cos \sigma^2 & -\sin \sigma^2  \\
                        0 & 0     &  \sin \sigma^2 & \cos \sigma^2
                              \end{array} \right),
\ \ \ S\in SO(4)
\end{equation}
applied to the vector $\vec{x_0}^T=(m_1,0,m_2,0)$ (lying in the $x_1x_3$ plane) 
simultaneously in the $x_1x_2$ and $x_3x_4$ planes. The similar block structure of the 
 rotation matrix preserves for higher N. 
Because of the arbitrariness of 
the  ${\bold m}$-curve other global symmetries of the membrane  surface, except of the 
rotational  $U(1)$ symmetry, are not assumed.

So, by construction the ansatz (\ref{anzats}) desribes one of the representatives 
of the family of the $U(1)$ invariant surfaces created by various  rotations 
of an closed curve in the $2N$ dimensional Euclidean space. 
Each of the members of the family has a $U(1)$ symmetry as its inherent global symmetry. 

The membrane world-volume metric $G_{\alpha \beta}$ corresponding to (\ref{anzats}) is
 \[
 G_{\alpha\beta}=diag(1- \dot{\bold{m}}^2, -\bold{m}^{' 2},-\bold{m}^2 ), 
\  \ \  \bold{m}:=(m_1,..,m_N).
 \]
The canonical momentum
$\boldsymbol{\pi}:=(\pi_1,..., \pi_N)$ conjugate  to
$\bold{m}$ and defined as
$\pi_a=\frac{\partial \mathcal{L}}{\partial \dot{m_a}}=
\vec{\mathcal{P}}\frac{\partial{\dot{\vec{x}}}}{\partial \dot{m_a}},
\ \ (a=1,2,.., N)$,
is presented in the explicit  form as
\begin{equation}\label{Imp}
\pi_a = \mathcal{P}_0 \dot{m}_a, \ \ \
\mathcal{P}_0 =\sqrt{\frac{\bold{m}^2\bold{m}^{' 2}}{1-\dot{\bold{m}}^2}} \ \ ,
\end{equation}
 where $\bold{m}^{'}$ and $\dot{\bold{m}}$ are partial derivatives
with respect to $\sigma^1$ and $t$, after using  (\ref{13}) and the relations: \,
$   \vec{x}^2=\bold{m}^2, \ \ \ \dot{\vec{x}}^2=\dot{\bold{m}}^2, \ \ \
\vec{x}^{' 2}=\bold{m}^{' 2},
\ \ \ g=\bold{m}^2\bold{m}^{' 2}  $.
 Then the Hamiltonian density (\ref{9}) and the constraints become
\[
\mathcal{H}_0=\mathcal{P}_0=\sqrt{\boldsymbol{\pi}^2+\bold{m}^2\bold{m}^{' 2}}, \ \ \
\dot{\mathcal{P}_0}=0,
\]
\begin{equation}\label{T1}
T:=T_1=\boldsymbol{\pi}\bold{m}^{'}=0.
\end{equation}
The corresponding Hamiltonian
equations of motion are transformed in Eqns.
\begin{equation} \label{hameq}
\dot{\bold{m}}=\{H_0,\bold{m}\}=\frac{1}{\mathcal{P}_0}\boldsymbol{\pi}, \ \ \ \
\dot{\boldsymbol{\pi}}=\{H_0,\boldsymbol{\pi}\}=
\left(\frac{\bold{m}^2 \bold{m}^{'}}{\mathcal{P}_0} \right)^{'},
\end{equation}
where the canonical  Poisson bracket and the Hamiltonian are defined by
\[
\{\pi_a(\sigma^1),m_b(\tilde{\sigma^1})\}=\delta_{ab}\delta(\sigma^1-\tilde{\sigma^1}), \ \ \ \
H_0=\int d\sigma^1 \sqrt{\boldsymbol{\pi}^2+\bold{m}^2\bold{m}^{' 2}}\ \  . \label{hameq1}
\]
The  Hamiltonian Eqns. (\ref{hameq}) yield the
following
system of PDE's
\begin{equation} \label{pde}
\ddot{\bold{m}}=\frac{1}{\mathcal{P}_0}\left( \frac{\bold{m}^2\bold{m}^{'} }{\mathcal{P}_0 }
\right)^{'}- \left( \frac{1}{\mathcal{P}_0} \bold{m}^{'} \right)^2 \bold{m} ,
\end{equation}
which coincides with equations (\ref{13}) after the substitution of (\ref{anzats}).

The gauge condition $\mathcal{P}_0=C$
transforms the constraint  $U$  to the following
\[
C^2\dot{\bold{m}}^2+\bold{m}^2\bold{m}^{' 2}-C^2=0,
\]
and the membrane dynamics is now completely determined by the
simplified  equations and constraints
\begin{eqnarray}
C^2\ddot{\bold{m}}=(\bold{m}^2\bold{m}^{'})^{'}-\bold{m}^{' 2}\bold{m} \label{eq11},\\
C^2\dot{\bold{m}}^2+\bold{m}^2\bold{m}^{' 2}-C^2=0, \ \ \ \ \dot{\bold{m}}\bold{m}^{'}=0. \label{eq2}
\end{eqnarray}
Eqns. (\ref{eq11}) and (\ref{eq2}) form the system of dynamical equations of
 the U(1)-invariant membrane (\ref{anzats})
and  their  study is the main goal of this paper.

To this end  we  note that Eqns. (\ref{eq11}) are invariant under the
scaling transformations: $\tau \to \tilde \tau=\frac{1}{ab} \tau$,
 $\sigma\to \tilde\sigma =\frac{1}{b} \sigma$, $\bold{m}\to
\tilde{\bold{m}}= a\bold{m}$, so that $\tilde{\bold{m}}(\tilde \tau, \tilde\sigma^1)=a\bold{m}(\tau,\sigma^1)$.
After performing the rescaling and fixing $a$ and $b$ by the condition
 $a^4b^2C^2=1$, the constraints (\ref{eq2}) transform into
\begin{equation}
C^2\dot{\bold{m}}^2+\bold{m}^2\bold{m}^{' 2}-1=0, 
\ \ \ \  \dot{\bold{m}}\bold{m}^{'}=0 \label{eq21}.
\end{equation}
Therefore making an additional change of the time variable into $\tau^*=\tau/C$ we find 
that the Eqns. (\ref{eq11},\ref{eq2}) transform into
\begin{eqnarray}
\ddot{\bold{m}}=(\bold{m}^2\bold{m}^{'})^{'}-\bold{m}^{' 2}\bold{m} \label{eq31},\\
\dot{\bold{m}}^2+\bold{m}^2\bold{m}^{' 2}-1=0, \ \ \ \ \dot{\bold{m}}\bold{m}^{'}=0. \label{eq32}
\end{eqnarray}
The above equations are  in agreement with \cite{JU1}, where they were derived by
a change of the variable $\sigma^1$. Moreover, it was observed  in \cite{JU1} that for $D=5$ 
(i.e. for $N=2$) the  constraints (\ref{eq32}) imply  Eqns. (\ref{eq31})  provided
 that $\bold{m}^{'}$ and $\dot{\bold{m}}$ are independent
\footnote {We observe that the static membrane characterized by Eqns. $\dot{\bold{m}}=0$,
 has to be treated
on a different footing.}. In the reduced phase space with $\mathcal{P}_0=C=1$
the Hamiltonian $H$ and the constraints take the form
\begin{eqnarray}
 H=\int d\sigma^1 \mathcal{H}= \int d\sigma^1 (\boldsymbol{\pi}^2+\bold{m}^2\bold{m}^{' 2}),
\label{Gam}\\
U=\boldsymbol{\pi}^2+\bold{m}^2\bold{m}^{' 2}-1=0, \ \ \ T=\boldsymbol{\pi}\bold{m}^{'}=0
\label{Svz}
\end{eqnarray}
with  the Hamiltonian density $\mathcal{H}|_U=\mathcal{P}_0=1$ on the constraint surface.

\subsection{General static solution}

The static membranes play a great role in M/string theory and it is important to understand
 the role
of the global $U(1)$ symmetry there. The obtained  answer is surprising: the dynamics of
static $U(1)$-invariant membrane
occurs to be exactly solvable  in any odd dimension $D=2N+1$, including $D=11$.

To prove this we  put  $\dot{\bold{m}}=0$ in  Eqns. (\ref{eq11}-\ref{eq2}) and obtain the equations
\begin{eqnarray}\label{sta1}
(\bold{m}^2\bold{m}^{'})^{'} -\bold{m}^{'2}\bold{m}=0,\\
\bold{m}^{'2}\bold{m}^2 -1 =0     \label{sta2}.
\end{eqnarray}
Using  (\ref{sta2}) we express  $\bold{m}^{'2}=1/\bold{m}^{2}$
and transform Eqns. (\ref{sta1})  to the  form
\begin{equation} \label{sta1'}
 \bold{m}^{2}(\bold{m}^2\bold{m}^{'})^{'} - \bold{m}=0,
\end{equation}
where the derivative $\partial_{\sigma^1}$ appears only in the
combination  $\bold{m}^{2}$$\partial_{\sigma^1}$.
This observation proposes a natural change of
the variable $\sigma^1$ by a new one $\xi(\sigma^1)$ defined by the differential relation
\begin{equation}\label{ksi}
d\xi= \frac{1}{\bold{m}^{2}(\sigma^1)}d\sigma^1,\ \ \ \ \ \ \  \bold{q}(\xi)=\bold{m}(\sigma^1).
\end{equation}
Using the variable $\xi$ one can present Eqns. (\ref{sta1}) and (\ref{sta2})
in the form
\begin{equation}
\partial_{\xi\xi}\bold{q}-\bold{q}=0, \ \ \ \ (\partial_{\xi}\bold{q})^2=\bold{q}^{2}  \label{lin1}
\end{equation}
of the $N$ linear differential equations with a simple nonlinear constraint.

The $\bold{q}$-system (\ref{lin1}) is equivalent to the original
$\bold{m}$-system (\ref{sta1}-\ref{sta2}) and the solution of one gives a solution of the other.
 The general solution  of the $\bold{q}$-system may be presented in the form
\begin{equation}\label{lnsol}
\bold{q}=\bold{C}_1 \cosh\xi+ \bold{C}_2 \sinh \xi, \ \ \ \ \ \
\bold{C}_1^2=\bold{C}_2^2=c^2,
\end{equation}
where $\bold{C}_1,\,\bold{C}_2$ are the integration constants.
 It proves the claimed exact solvability of the static $U(1)$-membrane equations.

The next step is to express the general solution (\ref{lnsol})
 in the  $\bold{m}$-representation. To this end
we need to integrate  Eq. (\ref{ksi}) to obtain $\xi$ as a function
of $\sigma^1$. The subsequent substitution of this
solution $\xi(\sigma^1)$ to (\ref{lnsol})
will create the desired general solution in the $\bold{m}$-representation.
Let us realize this program.

The integration of (\ref{ksi}) gives $\sigma^1$  as an explicit  function of $\xi$
\begin{equation}\label{implic}
\sinh 2\xi +\delta \cosh 2\xi=z(\sigma^1),
\ \ \
z:=\frac{2(\sigma^1 + \sigma_0^1)}{c^2},
\ \ \
\delta:=\frac{1}{c^2}(\bold{C}_1\bold{C}_2),
\end{equation}
where  $ \sigma_0^1$ and  $\delta$ are the integration constants, and
 $\delta \ \ (0\leq\delta\leq1)$ coincides with the cosine of
the angle between  $\bold{C}_1$ and $\bold{C}_2$.
On the other hand taking into account the relations following from
 (\ref{ksi}-\ref{lnsol})
\begin{equation}\label{scalprod}
(\bold{m}^4)^{'}/4=\bold{q}\partial_{\xi}\bold{q}=c^2(\sinh 2\xi +\delta \cosh 2\xi)
\end{equation}
and combining  (\ref{scalprod}) with (\ref{implic}) we  find
$\bold{m}^2$ as the explicit
 function of $\sigma^1$
\begin{equation}\label{msquar}
\bold{m}^2=c^2\sqrt{z^2 +4\gamma/c^4},
\end{equation}
where $\gamma$ is the integration constant.
The substitution of $\bold{m}^2(\sigma^1)$ (\ref{msquar}) in Eqn. (\ref{ksi})
and its integration in $\sigma^1$ gives the
desired  explicit presentation of $\xi(\sigma^1)$
\begin{equation}\label{ksisolut}
2\xi(\sigma^1)=  {\rm arcsinh}(\mu z) + \tilde\gamma,
\ \ \
\mu:= c^2/(2\sqrt{\gamma})=1-\delta^2,
\ \ 
\tilde\gamma=-{\rm arctanh\delta}
\end{equation}
The substitution of $\xi(\sigma^1)$ (\ref{ksisolut}) into the general
solution (\ref{lnsol}) yields the desired general solution in the $\bold{m}$-representation.

To discuss this general solution we note that because of the global
invariance of Eqns. (\ref{sta1}-\ref{sta2}) of the $\bold{m}$-system under the
scaling transformations
$\tilde\sigma^1 =b \sigma^1, \ \ \   \tilde{\bold{m}}= \sqrt{b}\bold{m}$ one can put
$\mu=1$ in (\ref{ksisolut}) without loss of generality.
This choice of $\mu$ is equivalent to $\delta=\tilde\gamma=(\bold{C}_1\bold{C}_2)=0$
and results in
\begin{equation}\label{ro}
2\xi(\sigma^1)= {\rm arcsinh}z,
\ \ \
\bold{m}(\sigma^1)^2= c^2\sqrt{z^2 + 1},
\ \ \
z:=2(\sigma^1 + \sigma_0^1)/c^2
\end{equation}
producing a simplified representation of the general solution of Eqns. (\ref{sta1}-\ref{sta2})
\begin{eqnarray}\label{sol2}
\bold{m}(\sigma^1)=\bold{C}_1 \sqrt{\frac{\cosh(2\xi) +1}{2}}+
\bold{C}_2 \sqrt{\frac{\cosh(2\xi) -1}{2}} = \\ \nonumber
\bold{C}_1\sqrt{\frac{\sqrt{1+z^2}+1}{2}}+
\bold{C}_2\sqrt{\frac{\sqrt{1+z^2}-1}{2}}.
\end{eqnarray}

If the  parameter $\sigma^1$ was bounded, e.g. $\sigma^1
\in [0,2\pi]$ - the case discussed  in more details in
Section 3 -  then the solution (\ref{sol2}) would be bounded too.
It corresponds to a membrane with the boundary, since the periodicity
condition $\bold{m}(0)=\bold{m}(2\pi)$ can not be
satisfied. On the other hand, if  we let $\sigma^1\in[0, \infty)$ then the general
 solution describes a
 $U(1)$-invariant  membrane without a boundary
\footnote{Since equations (\ref{eq11}) and
(\ref{eq2}) can be derived without referring to the Moser
theorem \cite{Mos}, the assumptions of the theorem (in particular the
compactness of the membrane) can be omitted allowing one to consider noncompact
membrane solutions.}.

The solution (\ref{sol2}) shows that the angle $\theta(\sigma^1)$
between $\bold{m}$ and $\bold{m}^{'}$
at any point $\sigma^1$ is given by the relation
\[
\bold{m}\bold{m}^{'}=\cos\theta= (1+z^{-2})^{-1/2},
\]
and goes to zero when $\sigma^1$ increases. The asymptotic
behavior of $\bold{m}(\sigma^1)$ when $\sigma^1 \rightarrow \infty $
is described by the vector function $\bold{m_{\infty}}(\sigma^1)$ is
\begin{equation}\label{asymt}
\bold{m(\sigma^1)} \rightarrow
\bold{m_{\infty}}=\bold{C}\sqrt{2\sigma^1},  \ \ \
\bold{C}:= (\bold{C}_1 + \bold{C}_2)/\sqrt2, \ \ \  \bold{C}^2=c^2.
\end{equation}
The comparison  of the product  $\bold{m_{\infty}}^2\bold{m_{\infty}^{'}}^2= c^4$
 with the constraint (\ref{sta2}) hints that $\bold{m_{\infty}}$ could be a special, exact
solution $\bold{m^{(pl)}}$ of Eqns. (\ref{sta1}-\ref{sta2})
provided that $\bold{C}$  satisfies the additional constraint  $\bold{C}^2=c^2=1$ that implies
\begin{equation}\label{plsol}
\bold{m^{(pl)}}(\sigma^1)= \pm\bold{n}\sqrt{2\sigma^1}, \ \ \
\end{equation}
where $\bold{n}$ is a unit constant vector with $N$ components.
The substitution of $\bold{m^{(pl)}}(\sigma^1)$ (\ref{plsol}) into
Eqns. (\ref{sta1}) proves that $\bold{m^{(pl)}}(\sigma^1)$ actually is
a special solution of (\ref{sta1}) characterized  by the additional
collinearity constraint
\begin{equation}\label{colin}
\bold{m^{(pl)}}(\sigma^1)\bold{m^{'(pl)}}(\sigma^1)=1.
\end{equation}
The constraint (\ref{colin}) means that the vectors
$\bold{m^{(pl)}}$ and $\bold{m^{'(pl)}}$ are parallel, but their
lengths $\rho^{(pl)}(\sigma^1)= \sqrt{2\sigma^1}$ and
$\rho^{'(pl)}(\sigma^1)= \frac{1}{\sqrt{2\sigma^1}}$ are inversely
proportional. So, the  gradient $\rho^{'(pl)}$ is equal to the
infinity at the end point $\sigma^1=0$ of the interval $[0,
\infty)$, and respectively to zero at $\sigma^1=\infty$.

For the first nontrivial case $N=2$, corresponding to $ D=5$,  the constraints (\ref{lnsol})
 for  the 2-vectors $\bold{C}_1$ and $\bold{C}_2$ of the $\bold{m}$-plane
 are  explicitly solved by introducing two arbitrary constants $C$ and $D$
\[
\bold{C}_1=(C,D), \ \ \ \ \bold{C}_2=(D,-C), \ \ \ C,D \in \mathbb{R}.
\]
In this  case the  solution (\ref{sol2}) yields a curve in the $\bold{q}$-plane
parametrized by $\xi$
\begin{eqnarray}\label{crvq}
 {q_{1}}^2 - {q_{2}}^2 = (C^2- D^2) + 2CD\sinh(2\xi)
\end{eqnarray}
which is obtained  from the  rectangular hyperbola: $\tilde q_{1}=c\,\cosh\xi,\ \ \tilde q_{2}=c\, \sinh\xi$
\begin{equation}\label{hyp1}
 {\tilde q_{1}}^2 - {\tilde q_{2}}^2 = c^2,\ \ \ c^2=C^2+D^2
\end{equation}
rotated in the $\bold{q}$-plane:
$q_{1}=\tilde q_{1}\cos\alpha  - \tilde q_{2}\sin\alpha, \ \ \ q_{2}= \tilde q_{1}\cos\alpha + \tilde q_{1}\sin\alpha$
with $\cos 2\alpha = (C^2-D^2)c^{-2}, \ \ \ \ \sin 2\alpha=-2CDc^{-2}$.
 Respectively,  the equation of the rotated hyperbola parametrized by $\sigma^1$
has the form
\begin{eqnarray}\label{crv2}
{m_{1}}^2 - {m_{2}}^2=C^2- D^2 +4CD(\sigma^1 + \sigma_0^1)/c^2.
\end{eqnarray}
On the other hand the asymptotic solution (\ref{plsol}) for $D=5$
\begin{equation}\label{plsol'}
\bold{m^{(pl)}}(\sigma^1)= \pm\bold{n}\sqrt{2\sigma^1}, \ \ \
\bold{n}=(\cos\psi_0, \sin\psi_0),
\end{equation}
parametrized by $\psi_0$, yields  a straight line in the
$\bold{m}$-plane presented by the equation
\begin{equation}\label{line}
m_{2}^{(pl)}=\lambda m_{1}^{(pl)}, \ \ \     \lambda = \tan \psi_{_0}.
\end{equation}
The asymptotic solution defines a membrane whose surface is a plane  going through the origin
of  the 4-dimensional  Euclidean  $\vec{x}$-space (\ref{anz})
\begin{eqnarray}\label{pleqn}
\vec{N}\vec{x}=0, \ \ \  \vec{N}=(-\sin\psi_0, -\sin\psi_0, \cos\psi_0, \cos\psi_0),
\end{eqnarray}
where $ \vec{x}^T=
\pm\sqrt{2\sigma^1}(\cos\psi_0\cos\sigma^2,\cos\psi_0\sin\sigma^2,\sin\psi_0\cos\sigma^2,\sin\psi_0\sin\sigma^2)$
is the world-volume vector of the $U(1)$ invariant membrane embedded in the five dimensional
Minkowski space-time.
Thus, we established that equations (\ref{sta1}),
(\ref{sta2}) are exactly solvable and their solutions (\ref{sol2}),
(\ref{plsol}) correspond to non-compact membranes having the above described  fixed shapes.

\subsection{The D=5 static solution in polar coordinates}

Here we study time dependent membrane solutions in D=5 using the
polar representation of the 2-dimensional vector $\bold{m}$ proposed
in \cite{JU1}. To this end we analyse how the polar representation
works in the discussed static case.

For $D=5$, the two-dimensional vector  $\bold{m}$ and
the constraint (\ref{sta2}) are
\begin{equation}\label{sanz}
\bold{m}=\rho(\sigma^1)\left(\cos \psi(\sigma^1),\sin \psi(\sigma^1) \right), \ \ \
\end{equation}
\begin{equation}\label{static1}
\rho^2(\rho^{' 2}+\psi^{' 2}\rho^2)=1 ,
\end{equation}
hence the  two components of a vector equation (\ref{sta1}) take  the form
\[
f_1 \rho \cos\psi+f_2\rho^2\sin\psi=0, \ \ \ \
\ \ \ \ \  f_1 \rho \sin\psi-f_2\rho^2\cos\psi=0,
\]
\[
f_1=2 \rho^2 \psi^{'2} - \rho^{'2} - \rho \rho^{''}, \ \ \ \   f_2=4\rho^{'}\psi^{'}+\rho\psi^{''}
\]
which generically imply that
\begin{equation}\label{static_5}
f_1=f_2=0.
\end{equation}
Because of the presence of the $\psi^{'}$ and $\psi^{''}$ in $f_2$, we
first consider the case when $\psi$ is a linear function $\psi=k\sigma^1+\psi_0$.

For $k=0$ the solution is $\rho=\pm\sqrt{2\sigma^1}$
and describes the non-compact membrane (\ref{plsol'}). The curve  $m_2 =F(m_1)$ in  the pane  $x_1x_3$
is a straight line and its  rotation (\ref{4rot})
yields a two parameter family of planes in $\mathbb{R}^4$ given by
\[
A(x_1-x_3\tan \psi_0) +B(x_2- x_4\tan \psi_0)=0 , \ \ \ A,B \in \mathbb{R}.
\]

For the case of  $k \ne 0$ we find that the solution of (\ref{static1}) is
\[
\rho(\sigma^1)=\pm \sqrt{\sin(2k\sigma^1+c)/k},
\]
where $k$ and $c$ are such that $\frac{1}{k}\sin(2k\sigma^1+c) \ge 0$.
 However, it turns out that this solution
does not satisfy (\ref{static_5}) or equivalently Eqns. (\ref{eq11}),
because
\[
\ddot{\bold{m}}-(\bold{m}^2\bold{m}^{'})^{'}+\bold{m}^{' 2}\bold{m} =
4k \rho(\sigma^1) (\sin(3k\sigma^1+c),-\cos(3k\sigma^1+c))\ne0.
\]
It shows that the ansatz $\psi=k\sigma^1+\psi_0$, $k\ne 0$, is a
solution of the constraint (\ref{static1}) but not of the
static-membrane equations.

Next, assuming that  $\rho \ne 0$ and that $\psi$ is not linear in $\sigma^1$, we find
from Eqn. (\ref{static_5})
 (i.e. from $f_2=0$) that $\rho^4\psi^{'}=1$ which substituted to $(\ref{static1})$
and (\ref{static_5}) (i.e. to $f_1=0$) gives
\begin{equation}
\rho=\pm\sqrt{2/b}\sqrt[4]{(a+b\sigma^1)^2+1}, \ \ \ \ \psi=\pm\arctan(a+b\sigma^1)/2+\phi, \ \ \
b>0.  \label{sol1}
\end{equation}
Note that the parameters $a$ and $b$ could be argued from the scaling invariance of the
equations: $\sigma^1 \to a+ b \sigma^1$, $\bold{m}\to \sqrt{b}\bold{m}$.

The curve $m_2=F(m_1)$ in the $x_1x_3$ plane  is now given by
\begin{equation}\label{curve}
m_1^2-m_2^2= 2\cos(2\phi)/b-2(a+b\sigma^1)\sin(2\phi)/b,
\end{equation}
 which is recognized as the hyperbola $m_1^2-m_2^2= \frac{2}{b}$ rotated in that plane
through the angle $\phi$ (cp. (\ref{sanz})).
Therefore, this  solution is equivalent to the non-compact membrane (\ref{crv2}) that
 can be seen by writing
\[
a=\frac{2\sigma_0}{C^2+D^2}, \ \ \ \ b=\frac{2}{C^2+D^2}, \ \ \
 \sin(2\phi)=-\frac{2CD}{C^2+D^2}, \ \ \ \ \cos(2\phi)=\frac{C^2-D^2}{C^2+D^2}
\]
and next identification of  $\phi$ with $\alpha$.
 The corresponding implicit equation of the membrane surface is given
by (cp. (\ref{anz}))
\[
 \left(x_1^2+x_2^2-x_3^2-x_4^2-2/b\cos(2\phi)\right)^2=b^2\sin(2\phi)^2
\left  ((x_1^2+x_2^2+x_3^2+x_4^2)^2 -4/b^2 \right),
\]
\[
 x_1x_4=x_2x_3.
\]

\section{Time dependent solutions}

Here we consider a class of solutions corresponding to  the following ansatz
\begin{equation}
\bold{m}=\rho(\xi)\left(\cos \psi(\eta),\sin \psi(\eta) \right), \ \ \ \xi=a_1\sigma +a_2 \tau, \ \ \
\eta=a_3\sigma + a_4 \tau. \label{ganz}
\end{equation}
In that case the  constraints (\ref{eq2}) become
\[
 a_1a_2 \rho_{\xi}^2+a_3a_4\psi_{\eta}^2=0,\ \ \
a_2^2 \rho_{\xi}^2+ a_4^2\rho^2\psi_{\eta}^2+\rho^2(a_1^2 \rho_{\xi}^2+ a_3^2\rho^2\psi_{\eta}^2) =1
\]
and one can see that if $a_i=0$ for at least one of $i$ then the solutions are either trivial
or correspond to
\[
\begin{array}{cccc}
a) & \bold{m} & = & \rho(\tau)\left(\cos \psi(\sigma^1),\sin \psi(\sigma^1) \right),  \\
b) & \bold{m} & = & \rho(\sigma^1)\left(\cos \psi(\tau),\sin \psi(\tau)\right),  \\
c) & \bold{m} & = & \rho(\sigma^1)\left(\cos \psi(\sigma^1),\sin \psi(\sigma^1)\right), \\
d) & \bold{m}&=&\rho(\tau)\left(\cos \psi(\tau),\sin \psi(\tau) \right),
\end{array}
\]
from which one can exclude the previously studied static case c).
The purely time dependent case d) yields a degenerate world-volume
metric, as $\det g=0$ everywhere, and describes the closed (in
$\sigma^2$) $U(1)$ invariant null string moving with the velocity of
light ($\dot{\bold{m}}^2=1$).

On the other hand if $a_i \ne 0$ for all $i$ then the first constraint is solved by
\[
\rho=\rho_0 e^{\pm \frac{a_3a_4}{a_1a_2}\psi_0^2 \xi },
\ \ \ \psi=\psi_0 \eta + \psi_1, \ \ \ a_1a_2a_3a_4>0
\]
in contradiction with the second constraint hence we can narrow the
discussion to a) and b) that correspond to compact membranes.

Since there are two possible topologies corresponding to a U(1) invariant,
compact, orientable membrane surface $\Sigma$ without boundaries - the sphere  $S^2$ with the genus
 $\bold{g}=0$ and the torus  $T^2$ with  $\bold{g}=1$ -
we shall take advantage of using the Gauss-Bonnet theorem
\begin{equation}\label{tgb}
\frac{1}{2\pi}\int_{\Sigma}d^2\sigma\sqrt{g} K=\chi(\Sigma)=2-2\bold{g},
\end{equation}
where $\chi(\Sigma)$ is  the
 Euler characteristic and $K$ is the Gauss curvature
 \begin{equation}  \label{K}
K=-\frac{1}{2}\frac{1}{\sqrt{\bold{m}^2\bold{m}^{'2}}}\left(
\frac{1}{\sqrt{\bold{m}^2\bold{m}^{'2}}} \bold{m}^{2'}  \right)^{'}
\end{equation}
given by $\vec{x}$ (\ref{anz}).
The integration range of $\sigma^1$ is $[0,\pi]$ or
$[0,2\pi]$ for $S^2$ or $T^2$ respectively.
The integration  over $\sigma^2 \in [0,2\pi]$
in  (\ref{tgb}) gives the equation
\begin{equation}\label{gb}
\int d\sigma^1
\left( \frac{1}{\sqrt{\bold{m}^2\bold{m}^{'2}}} \bold{m}^{2'}
\right)^{'} = 4\bold{g}-4
\end{equation}
with the  vanishing  integrand for the cases  a) and b). It  results in $\bold{g}=1$ and proves  that the a) and b)  cases  describe the surface $\Sigma$ which is a torus.

\subsection{Contracting  tori}

Let us at first  consider the ansatz a). The constraints (\ref{eq2}) are satisfied provided
$\dot{\rho}^2+\psi^{' 2}\rho^4=1 $ which implies that $\psi^{' 2}=\omega^2=const.$ and
\[
F\left(\arccos(\sqrt{\omega}\rho);1/\sqrt{2}\right)=\mp\sqrt{2\omega}(\tau+\tau_0),
\ \ \ \tau_0\in \mathbb{R},
\]
where $F(\phi;k)$ is the elliptic integral of the first kind.
It follows that for fixed $\tau$,
$\bold{m}(\sigma^1,\tau)$ is bounded, so that the membrane is compact
and in view of $K=0$ corresponds to a torus.  That
implies that $\omega=n$ $(n\in \mathbb{Z})$ due to the periodicity
condition $\bold{m}(\sigma^1,\tau)=\bold{m}(\sigma^1+2\pi,\tau)$.
Thus the solution reads
\[
\rho(\tau;n)=\frac{1}{\sqrt{n}}cn\left(\sqrt{2n}
(\tau+\tau_0),\frac{1}{\sqrt{2}}\right), \ \ \ \ \psi(\sigma^1)=n\sigma^1
+\psi_0, \ \ \ \tau_0,\psi_0\in\mathbb{R},  \ \ \ \ n \in \mathbb{Z},
\]
where $cn(\phi,k)$ is the Jacobi elliptic cosine function.
If the initial data are such that
with $\dot{\rho}(\tau_0)>0$ then
the solution describes an expanding torus which at
some point reaches the maximal size $\rho_{max}=1/\sqrt{n}$
and then shrinks to a point after a finite time
$\bold{ K}(1/\sqrt{2})/\sqrt{2n}$
(where $\bold{ K}(1/\sqrt{2})=1,8451$ is the quarter period of  the elliptic
cosine function).

The dynamical equation (\ref{eq11}) is automatically satisfied because $\bold{m}^{'}$
and $\dot{\bold{m}}$
are independent in this case.
The implicit presentation of time dependent surface $\Sigma(\tau)$
corresponding to the solution is given by
\[
 x_1^2+x_2^2+x_3^2+x_4^2=\frac{1}{n}cn\left(\sqrt{2n}(\tau+\tau_0),\frac{1}{\sqrt{2}}\right)^2,
\ \  x_1x_4=x_2x_3.
\]

\subsection{ Spinning torus}

Finally we study the torus ansatz b) characterized by $\bold{m}^2=\rho^2$ and
$\bold{m}^{' 2}=\rho^{' 2}$ that results in $K=0$
and the following solutions of the constraints (\ref{eq2})
\[
\rho(\sigma^1)^2=-\left(\omega \sigma^1\pm
\tilde{B}/\omega\right)^2+\omega^{-2},  \ \ \ \psi(\tau)=\omega \tau
+\psi_0,   \ \ \ \ \omega,\psi_0,\tilde{B} \in \mathbb{R}.
\]
Note that $\sigma^1$ is bounded since $\rho(\sigma^1)^2>0$
and the membrane is  compact similarly to the a) case.
The periodicity condition
$\rho(0)^2=\rho(2\pi)^2$  fixes the integration constant $\tilde{B}:
\omega^2\pi=\mp \tilde{B}$
representing the solution in the form
\begin{equation}
\rho(\sigma^1)^2=\omega^{-2}-(\pi-\sigma^1)^2\omega^2 , \ \ \ \psi(\tau)=\omega \tau +\psi_0  \label{spanz}
\end{equation}
describing a rotating torus with the preserved shape.
Here $\bold{m}^{'}$ and $\dot{\bold{m}}$
stay independent hence the ansatz b) with (\ref{spanz}) is the solution of
(\ref{eq11}). The function $\rho(\sigma^1)^2$ has a maximum at
$\sigma^1=\pi$ with $\rho(\pi)=1/\omega$ and a minimum at
$\sigma^1=0$ with
\[
\rho(0)=\rho(2\pi)=\frac{1}{\omega}\sqrt{1-\pi^2\omega^4}
\]
which shows  that  $\omega$ has the  upper limit
$\omega\le \omega_{max}=1/\sqrt{\pi}$.
Restoring the physical units ($[\Omega]=L^{-1}$) for  $\omega$ with help of
the membrane tension $T$,
one can  present the critical frequency in the form
\[
\Omega_{max}=T^{1/3}\omega_{max}=T^{1/3}/\sqrt{\pi} \quad .
\]
This time the implicit description of $\Sigma(t)$ is given by
\[
x_1^2+x_2^2+x_3^2+x_4^2=\omega^{-2}-\left(\pi-\arctan x_2/x_1 \right)^2\omega^2,
 \ \  x_1x_4=x_2x_3.
\]

\section{Summary}

The $U(1)$ globally symmetric membranes realizing the idea of
 reduction of the 3-dim nonlinear problem to a 2-dim string-like one were studied.
 Such type of reductions could sharpen the role of the methods developed
in the theory of 2D chiral models \cite{FaTa} for the membrane quantization problem.
To this end an attempt to integrate the  $U(1)$ membrane equations was undertaken here.
We established the exact solvability of these  nonlinear equations
for the static membranes in the odd dimensions $D= 5, 7, 9, 11,.., 2N+1$
of the Minkowski space, and found their general solution.
The general solution  for $D= 5$ includes
the infinitely extended membranes whose surface is a 2-dimensional plane.
We also found the time  dependent solutions  for
$D=5$ that show a deep connection of the elliptic functions with
 the membrane  dynamics (see also \cite{hoppe7}).
These functions encode peculiarities of the
hidden structure of the  membrane dynamics different from the string dynamics.
We hope that the solutions presented here will prove useful in further study of the integrability
of the  membrane dynamics along the line studied in \cite{hoppe2,hoppe7}.

\noindent{\bf Acknowledgments}

We are grateful to J. Hoppe for valuable collaboration, kind support
and critical remarks. AZ would like to thank  J.
{\AA}man, I. Bengtsson, R. Myers, N. Pidokrajt and N. Turok for useful discussions,
 and Fysikum at
Stockholm University, Department of Mathematics of Royal Inst. of
Technology KTH Stockholm, Nordic Inst. for Theoretical Physics
Nordita and Perimeter Inst. for Theoretical Physics for kind hospitality.
 MT would like to thank Albert Einstein Institute for hospitality.
This research was supported in part by
Knut and Alice Wallenberg Foundation, Nordita, Marie Curie Research Training Network
ENIGMA (contract MRNT-CT-2004-5652)  and  Perimeter Institute for Theoretical Physics .


\begin{thebibliography}{99}


\bibitem{M1} P. K. Townsend, \textit{The eleven-dimensional supermembrane revisited},
Phys. Lett. B350 (1995) 184-187
;
E. Witten, \textit{ String Theory Dynamics In Various Dimensions}, Nucl. Phys. B443 (1995)
85-126.

\bibitem{tucker}
R. W. Tucker, \textit{Extended Particles and the Exterior Calculus}, Lectures given at the Rutherford Laboratory, Feb 1976.

\bibitem{hoppe1}
J. Hoppe, H. Nicolai, \textit{Relativistic minimal surfaces}, Phys. Lett. B196 (1987) 451.

\bibitem{Z_0}
A. A. Zheltukhin,  \textit{ A Hamiltonian of null strings:
An invariant action of null (super)membranes},
 Sov. J.  Nucl.  Phys. 48 (1988) 375-379.

\bibitem{BZ_0}
I. A. Bandos and A. A. Zheltukhin,
 \textit{Null super p-branes quantum theory in four-dimensional space-time},
 Fortsch. Phys. 41  (1993) 619-676.

\bibitem{hoppe2}
M. Bordemann, J. Hoppe, \textit{The Dynamics of Relativistic
Membranes I: Reduction to 2-dimensional Fluid Dynamics}, Phys. Lett.
B317 (1993) 315-320, {\tt arXiv:hep-th/9307036};
\textit{The Dynamics of Relativistic
Membranes II: Nonlinear Waves and Covariantly Reduced Membrane
Equations}, Phys. Lett. B325 (1994) 359-365, {\tt arXiv:hep-th/9309025}.

\bibitem{hoppe4}
 J. Hoppe, \textit{Some Classical Solutions of Relativistic Membrane
Equations in 4 Space-Time Dimensions}, Phys. Lett. B329 (1994).
10-14;
\textit{Canonical 3+1 description of relativistic  membranes},{\tt arXive:hep-th/9407103v2}.

\bibitem{UZ}
 D. V. Uvarov and A. A. Zheltukhin,
\textit{Exactly solvable p-brane model with extra supersymmetry},  Phys. Lett. B545 (2002) 183-189.

\bibitem{hoppe6}
J. Arnlind, J. Hoppe, S. Theisen, \textit{Spinning membranes}, Phys, Lett, B 599 (2004) 118-128.

\bibitem{JU1}
J. Hoppe, \textit{U(1) invariant  Membranes and Singularity Formation}, {\tt arXiv:0805.4738v1}.


\bibitem{Reb}
C. Rebbi, Phys. Rep. V12 N1 (1974) p. 1-73.


\bibitem{Mos}
J. Moser, \textit{On the volume elements of a manifold},
Trans. Amer. Math. Soc., V120 (1965) 287.

\bibitem{FaTa}
L. D. Faddeev, L. A. Takhtajan,  \textit{Hamiltonian Methods in the
Theory of Solitons, Classics in Mathematics}, Berlin, Springer, (1987).

\bibitem{hoppe7} J. Hoppe,
\textit{Conservation Laws and Formation of Singularities in Relativistic Theories
of Extended Objects},{\tt arXiv:hep-th/9503069}.

\end{thebibliography}
\end{document}